%% file: main.tex
\documentclass[aps,prd,
floatfix,preprintnumbers,superscriptaddress,nofootinbib,nameinlink,capitalise]{revtex4-2}

\usepackage[left=2.54cm, right=2.54cm, top=2.54cm, bottom=2.54cm]{geometry}
\usepackage{palatino}
\usepackage{amsmath, amssymb}
\usepackage{graphicx}
\usepackage{array, booktabs, multirow}
\usepackage{float}
\usepackage{float} 
\usepackage{hyperref}
\usepackage{subfig}
\usepackage{nameref}      
\usepackage{cleveref}
\usepackage{placeins}
\usepackage{verbatim}
\usepackage{textgreek}
\usepackage{setspace}
\usepackage[table]{xcolor}   
\usepackage{booktabs}        

\usepackage[justification=raggedright,singlelinecheck=false]{caption}
\usepackage{color}
\newcommand{\bmat}{\left(\begin{array}}
\newcommand{\emat}{\end{array}\right)}
\newcommand{\be}{\begin{equation}}
\newcommand{\ee}{\end{equation}}
\newcommand{\bea}{\begin{eqnarray}}
\newcommand{\eea}{\end{eqnarray}}

\usepackage{graphicx}
\usepackage{subcaption}

\begin{document}

\title{Radiatively Corrected Hybrid Inflation: Parameter Scans and Machine Learning with ACT and Future CMB Experiments}

\input{authors.tex}
\noaffiliation


\vspace{1em}
\begin{abstract}

We investigate a realistic non-supersymmetric hybrid inflation model incorporating right-handed neutrinos and assess its viability in light of recent cosmological observations. At tree level, the inflaton potential yields a blue-tilted scalar spectrum, which is disfavored by current data from Planck and ACT that instead support a red tilt. We show that including one-loop quantum corrections, arising from generic couplings required for reheating, significantly modifies the potential, flattening it at large field values. This leads to a red-tilted spectral index ($n_s < 1$) and a suppressed tensor-to-scalar ratio $r$, both consistent with observational constraints. To ensure theoretical control, we focus on sub-Planckian field values, where the effective field theory description remains valid. The coupling of the inflaton to right-handed neutrinos naturally facilitates efficient reheating and enables the generation of the baryon asymmetry via non-thermal leptogenesis. We further explore the model’s parameter space using a multi-output random forest classifier, achieving prediction accuracies in the range of $87.5\%$ to $98.9\%$. Our analysis shows that approximately $15\%$ of the parameter space satisfies at least one current experimental constraint, underscoring the essential role of quantum corrections in reconciling particle physics models with precision cosmology, and highlighting the effectiveness of machine learning techniques in probing complex theoretical frameworks.

\end{abstract}

\maketitle

\section{Introduction}

One of the pillars of modern cosmology is the inflationary paradigm which has provided a solid justification for the extensive homogeneity, isotropy and flatness of the universe as well as being the genetic antecedent of the primordial density fluctuations that give raise to the cosmic structures that we observe in the current world \cite{Guth:1980zm,Linde:1981mu,Albrecht:1982wi}. Among the diverse landscape of inflationary models, hybrid inflation \cite{Linde:1993cn} introduced a particularly elegant mechanism for ending the inflationary epoch. In this framework, a "waterfall" phase transition, triggered when the inflaton field rolls past a critical point, provides a swift and graceful exit a feature that makes hybrid inflation exceptionally appealing for embedding within high-energy particle physics theories, such as Grand Unified Theories (GUTs) \cite{Senoguz:2003zw}.

The hybrid inflation scenario proposed by Linde ~\cite{Linde:1993cn} includes a simple potential along the inflationary  trajectory, in which a constant vacuum energy $V_{0}$ dominates, supplemented by a quadratic term for the inflaton,
\begin{equation}
    V(\phi) = V_{0} + \frac{1}{2} m^{2}\phi^{2}.
\end{equation}
Despite its theoretical appeal, the tree-level predictions of this model are now in significant tension with precision cosmological data. In particular, it typically yields a blue-tilted spectrum of scalar perturbations ($n_{s} \geq 1$), disfavored by the latest results from Planck and the Atacama Cosmology Telescope, both of which favor a red-tilted spectrum ($n_{s} < 1$)~\cite{Planck:2018jri, ACT:2025tim}. A compelling resolution to this tension arises from quantum field theory itself. The inflaton must couple to other fields to facilitate the reheating of the universe after inflation. These couplings inevitably induce radiative corrections, which can dramatically alter the inflationary dynamics. Incorporating the celebrated Coleman-Weinberg one-loop corrections \cite{Coleman:1973jx} leads to a modified potential:
\begin{equation}
    V(\phi) = V_{0} + \frac{1}{2} m^{2}\phi^{2} + A\,\phi^{4}\ln\!\left(\frac{\phi}{\mu}\right),
\end{equation}
where the sign and magnitude of the parameter $A$ are determined by the couplings of the inflaton to bosonic and fermionic fields. In the regime where fermionic loop contributions dominate ($A < 0$), the potential becomes flatter at large field values, which naturally leads to a red-tilted scalar spectral index $n_{s}$ and a suppressed tensor-to-scalar ratio $r$~\cite{Senoguz:2008nok,Rehman:2009wv}.

In this work, we undertake a detailed investigation of a nonsupersymmetric hybrid inflation scenario, incorporating a fully consistent treatment of the radiative corrections that reshape the inflaton potential. Our analysis substantially extends earlier studies by performing a wide-ranging numerical exploration of the model parameter space and confronting the resulting predictions with the most stringent combined cosmological datasets currently available, including P-ACT-SPT, P-ACT-LB-BK18\cite{ACT:2025tim}. \footnote{In this context, considerable recent work has been devoted to examining cold inflationary models, within both supersymmetric and non-supersymmetric frameworks, confronted with the observational constraints reported by the Atacama Cosmology Telescope (ACT); see Refs. \cite{Rehman:2025fja, Kallosh:2025rni,Ahmed:2025crx, Pallis:2025epn, Ahmed:2025eip, Pallis:2026cyz, Ahmed:2025sfm, Ahmed:2025rrg,Zharov:2025zjg,Ketov:2025cqg, Ellis:2025zrf, Leontaris:2025hly,McDonald:2025tfp,Pallis:2025nrv,Pallis:2025gii,Gao:2025onc,Pallis:2025vxo,Liu:2025qca,Odintsov:2025eiv,Gialamas:2025kef,Okada:2025nyd,Modak:2025bjv,Ellis:2026ceb,Pallis:2026qci, Pallis:2026cyz}}
.

A central result of this study is the establishment of full post-inflationary consistency within the same theoretical framework. We compute the reheating temperature in a self-consistent manner and demonstrate that the fermionic couplings to right-handed neutrinos, which generate the requisite negative radiative contribution to the inflaton potential, simultaneously:

\begin{itemize}
    \item provide a viable and efficient reheating mechanism through inflaton decay, and
    \item naturally yield the observed baryon asymmetry via non-thermal leptogenesis~\cite{Fukugita:1986hr,Lazarides:1990huy,Lazarides:1996dv},
\end{itemize}

thereby satisfying the empirical requirement on the baryon-to-entropy ratio,
\begin{equation}
    \frac{n_{B}}{s} \simeq 8.7 \times 10^{-11}.
\end{equation}

Moreover, the inclusion of high-precision small-scale measurements from ACT and SPT significantly tightens the constraints on the scalar spectral index and tensor-to-scalar ratio, providing a more definitive assessment of the model's viability in light of current observational data.

Furthermore, we exploit modern machine-learning techniques to efficiently probe the high-dimensional parameter space of the model. In particular, we employ  Multi-Output Random Forest classifier \cite{rF} to systematically quantify the regions of phenomenological viability under current and prospective observational constraints. This analysis reveals that the radiative correction parameter $A$ plays the dominant role in determining compatibility with data, effectively controlling the flattening of the potential and the resulting inflationary observables.

Our results demonstrate that the radiatively corrected hybrid inflation framework remains not only viable but also highly predictive. The model can be realized with sub-Planckian field excursions, preserves theoretical control within the effective field theory regime, and predicts a tensor-to-scalar ratio within the anticipated sensitivity of forthcoming CMB polarization missions such as LiteBIRD and CMB-S4~\cite{LiteBIRD:2022cnt, CMB-S4:2016ple}. Taken together, these findings establish a coherent and phenomenologically successful framework that links a well-motivated inflationary dynamics to a consistent post-inflationary thermal history grounded in observational evidence.

The paper is organized as follows. In Section~II, we introduce the radiative corrections to the inflaton potential, outlining the theoretical framework and motivating the inclusion of loop-level effects. Section~III focuses on reheating and non-thermal leptogenesis, where we analyze how inflaton decay can successfully generate the observed baryon asymmetry of the Universe. In Section~IV, we perform a comprehensive numerical analysis to identify regions of the parameter space that are consistent with the latest cosmological observations from the \textit{Planck} and ACT collaborations. Section~V is devoted to a detailed {machine learning analysis of cosmological parameter compatibility, where we assess the viability of the model across different experimental configurations and confirm key theoretical insights through data-driven methods. Finally, in Section~VI, we summarize our findings and discuss their broader implications, including potential observational signatures that could test the validity of the proposed framework.

\section{Inflationary Dynamics in Radiatively Corrected Hybrid Models}
\label{sec4}
In a general renormalizable theory, the Lagrangian contains many possible interaction terms up to mass dimension four. 
The full dynamics, including kinetic terms, scalar potential, and Yukawa interactions involving the real scalars $\phi$ and $\chi$ and right-handed Majorana fermion $N = N^c$, are described by the Lagrangian
\begin{eqnarray}
\mathcal{L} & = & \frac{1}{2} \partial^\mu \phi \, \partial_\mu \phi + \frac{1}{2} \partial^\mu \chi \, \partial_\mu \chi + \frac{i}{2}  \overline{N} \gamma^\mu \partial_\mu N \nonumber\\
&& - \kappa^2 \left( M^2 - \frac{\chi^2}{4} \right)^2 - \frac{m^2}{2} \phi^2 - \frac{g^2}{4} \chi^2 \phi^2 - \frac{\lambda_{\phi}}{4!} \phi^4 \nonumber\\
&& - \frac{1}{2} y_\phi \, \phi \, \overline{N} N - \frac{1}{2} \frac{\lambda_{\chi}}{m_{\rm Pl}} \, \chi^2 \, \overline{N} N .
\label{eq:full_lagrangian}
\end{eqnarray}
Here, $\kappa$, $g$, $\lambda_{\phi}$, $y_{\phi}$, and $\lambda_{\chi}$ are dimensionless couplings, $M$ is the symmetry-breaking scale associated with $\chi$, $m$ is the inflaton mass parameter, and the term $-\frac{\lambda_{\phi}}{4!} \phi^4$ represents the quartic self‑interaction of the inflaton. The term $\frac{g^2}{4} \phi^2 \chi^2$ provides the coupling between the inflaton $\phi$ and the waterfall field $\chi$.

The above form of the Lagrangian can be obtained by imposing some discrete symmetries. Specifically, a discrete symmetry  $\mathbb{Z}_4^{\phi} \times \mathbb{Z}_2^{\chi}$  can be imposed to independently constrain the fields $\phi$, $\chi$, and $N$, acting as
\begin{equation}
\mathbb{Z}_4^{\phi }: \quad \phi \to -\phi,  \quad N \to i N, \qquad 
\mathbb{Z}_2^{\chi}: \quad \chi \to -\chi.
\end{equation}
These symmetries forbid all odd-power scalar operators, including $\phi$, $\chi$, $\phi^3$, $\chi^3$, and mixed terms such as $\phi\chi$, $\phi^2\chi$, $\phi\chi^2$, $\phi^3\chi$, and $\phi\chi^3$, while allowing only even-power combinations like $\phi^2$, $\chi^2$, $\phi^4$, $\chi^4$, and $\phi^2\chi^2$.
In the fermion sector, the Yukawa interaction $\phi \, \overline{N}N$ remains invariant under these symmetries, whereas the coupling $\chi \, \overline{N}N$ is forbidden. Likewise, a bare Majorana mass term $m_N \overline{N}N$ is not allowed. As a result, the scalar potential naturally acquires the desired hybrid inflation form with minimally required realistic interactions. 

Instead of imposing discrete symmetries, one may alternatively consider scenarios in which the waterfall field transforms non-trivially under a gauge symmetry, such as a $U(1)$ extension of the Standard Model \cite{Okada:2011en} or a simple grand unified gauge group like $SU(5)$ or $SO(10)$ \cite{Rehman:2008qs}. In contrast, the inflaton is typically required to remain a gauge singlet to suppress large radiative corrections during inflation, which could otherwise spoil the slow-roll conditions.

The last term, characterized by the dimensionless coupling $\lambda_{\chi}$, can arise from higher-dimensional operators and may be interpreted as originating from a mild breaking of the discrete symmetries at the nonrenormalizable level. 
Such a breaking, potentially originating from an ultraviolet-complete theory, is also crucial for avoiding the domain wall problem that could otherwise result from the spontaneous breaking of the $\mathbb{Z}_2^{\chi}$ symmetry after inflation.
After symmetry breaking, when $\chi$ develops a vacuum expectation value (vev), $\langle \chi \rangle = 2M$, the right-handed neutrinos acquire Majorana masses of order $\lambda_{\chi} M^2 / m_{\rm Pl}$. This naturally provides the required suppression scale for implementing the seesaw mechanism and explaining the smallness of the observed neutrino masses. In contrast, the scalar field $\phi$ does not acquire a vev. The Yukawa interactions in the last line play an important role in reheating and leptogenesis, as will be discussed later.

The scalar potential relevant for hybrid inflation can be read off as follows:
\begin{eqnarray}
V(\phi,\chi) = \kappa^2 \left(M^2 - \frac{\chi^2}{4}\right)^2 + \frac{m^2}{2}\phi^2 + \frac{g^2}{4}\chi^2\phi^2 + \frac{\lambda_{\phi}}{4!}\phi^4.
\label{eq:potential}
\end{eqnarray}
A similar form of the potential was originally introduced in Linde’s hybrid inflation model \cite{Linde:1993cn}, albeit without the final quartic term. However, as discussed earlier, the tree-level predictions of this minimal setup are not consistent with current observational data, even in the presence of the final quartic term. As we demonstrate below, radiative corrections to the potential, arising from the well-motivated coupling between the inflaton and right-handed neutrinos, play a crucial role in bringing the model’s predictions into agreement with the latest experimental results.

The inflationary trajectory lies along the valley $\chi = 0$, where the potential simplifies to:
\begin{eqnarray}
    V(\phi) = V_0 + \frac{1}{2} m^2 \phi^2  + \frac{\lambda_{\phi}}{4!} \phi^4, \qquad \text{with} \quad V_0 \equiv \kappa^2 M^4.
\end{eqnarray}
Inflation is primarily driven by the constant vacuum energy $V_0$, with the $\phi$ terms providing a gentle slope. The stability of the $\chi = 0$ valley depends on its effective mass:
\begin{eqnarray}
   m_\chi^2(\phi) = \left. \frac{\partial^2 V}{\partial \chi^2} \right|_{\chi=0} = - \kappa^2 M^2 + \frac{g^2}{2} \phi^2 \,. 
\end{eqnarray}
This mass squared is positive for $\phi > \phi_c$, where the critical point is
\begin{equation}
\phi_c \equiv \frac{ \sqrt{2} \, \kappa \, M}{g}.
\end{equation}

The inflaton potential now also includes a quartic self-interaction term, $\frac{\lambda_\phi}{4!}\phi^4$, which, although subdominant compared to the vacuum energy $V_0 = \kappa^2 M^4$, can still influence the inflationary dynamics. In the false vacuum dominated regime, defined by $V_0 \gg \frac{1}{2}m^2\phi^2, \, \frac{\lambda_\phi}{4!}\phi^4$, the vacuum energy drives inflation, while both the quadratic and quartic terms contribute to the slow-roll evolution.
As the inflaton approaches the critical value $\phi_c^2 = 2\kappa^2 M^2/g^2$, the waterfall field becomes tachyonic, triggering a rapid phase transition that ends inflation. This graceful exit mechanism remains intact irrespective of the precise shape of the inflaton potential near $\phi_c$.

In the false vacuum dominated limit, the slow-roll predictions receive contributions from both mass and quartic terms. The spectral index and tensor-to-scalar ratio are approximately given by
\begin{equation}
n_s \simeq 1 + 2 m_{\rm Pl}^2 \frac{m^2 + \frac{\lambda_\phi}{2}\phi^2}{V_0},
\qquad
r \simeq 8 m_{\rm Pl}^2 \frac{\left(m^2 \phi + \frac{\lambda_\phi}{6}\phi^3\right)^2}{V_0^2}.    
\end{equation}
These expressions explicitly show that, even when subdominant in the potential, the quartic term can leave a non-negligible imprint on inflationary observables.
The spectral index, however, is still blue-tilted ($n_s \ge 1$), and the tensor-to-scalar ratio is typically very small. The blue tilt is the primary source of tension with current  P-ACT-SPT, P-ACT-LB-BK18\cite{ACT:2025tim} observational data.  To reconcile the predictions of the model with observations, we include radiative corrections arising from couplings of the inflaton to additional fields. These quantum corrections, calculable via the Coleman-Weinberg mechanism, modify the inflaton potential and can generate a slight red tilt ($n_s < 1$) and an appropriate amplitude of scalar perturbations, making the predictions consistent with current data. 

Specifically, we introduce three right-handed neutrinos $N_i$ (with $i=1,2,3$), which are singlets under the Standard Model gauge group. Their inclusion is well-motivated by the seesaw mechanism for neutrino mass generation and by leptogenesis. The relevant part of the Lagrangian contains Yukawa interactions:
\begin{eqnarray}
\mathcal{L} \supset -\frac{1}{2} y_\phi^{ij} \, \phi \, \overline{N}_i N_j ,    
\end{eqnarray}
where $y_\phi$ denotes the Yukawa coupling matrix. This interaction induces field-dependent masses for the right-handed neutrinos,
\begin{equation}
M_R(\phi) \approx y_\phi \, \phi ,
\end{equation}
during inflation (with $\chi=0$). 
In the inflationary regime, the one-loop Coleman–Weinberg potential receives contributions from both the scalar field $\chi$ and the right-handed neutrinos $N_i$. Under the assumptions $y_\phi^2 \gg \lambda_\phi$ and $y_\phi \phi \gg m$, the sum of the loop contributions yields
\begin{eqnarray} \label{CWP}
   V_{\rm 1-loop}(\phi) \simeq \frac{1}{64 \pi^2} \left[ \frac{g^4}{4} (\phi^2 - \phi_c^2)^2 \ln \left( \frac{g^2 (\phi^2 - \phi_c^2)}{2 \mu^2} \right) - 2 N_N (y_\phi \phi)^4 \ln \left( \frac{(y_\phi \phi)^2}{\mu^2} \right) \right], 
\end{eqnarray}
where $N_N = 3$ denotes the number of right-handed neutrino species and $\mu$ is the renormalization scale. The first (bosonic) term arises from $\chi$-field loops, while the second (fermionic) term originates from neutrino loops.

The renormalization group (RG) improved scalar potential can be written as
\begin{equation}
V(\phi) \supset \frac{1}{2} m^2 (t) \, G(t)^2 \phi^2  +  \frac{1}{4!} \lambda_{\phi}(t) \, G(t)^4 \, \phi^4,
\label{potrgi}
\end{equation}
where $t=\ln(\phi/\mu)$, and 
\begin{equation}
G(t) = \exp\left(- \int_0^t \frac{\gamma_\phi(t')}{1 + \gamma_\phi(t')} dt'\right) \approx 1. 
\end{equation}
Here, $\gamma_\phi = -\frac{3 y_\phi^2}{16 \pi^2} \ll 1$ is the anomalous dimension of the inflaton field.

The renormalization group equations (RGEs) for the relevant couplings are given by
\bea
\frac{d\lambda_{\phi}}{dt} &=& \frac{1}{(4\pi)^2} \left( 3 \lambda_\phi^2 + 3 g^4 + 4 \lambda_\phi \text{Tr} [y_\phi^{\dagger} y_\phi]  - 24 \, \text{Tr} [ (y_\phi^{\dagger} y_\phi)^2 ] \right), \label{eq:rge_lambda} \\[10pt]
\frac{d g }{dt} &=& \frac{ g }{(4\pi)^2} \left( 2 g^2 + \frac{1}{2} \lambda_\phi + \frac{3}{4} \kappa^2 +  \text{Tr} [y_\phi^{\dagger} y_\phi]   \right), \\[10pt]
\frac{d \kappa^2 }{dt} &=& \frac{1}{(4\pi)^2} \left( \frac{9}{2} \kappa^4 + 2 g^4  \right),  \\[10pt]  
\frac{d y_{\phi}}{dt} &=& \frac{y_{\phi}}{(4\pi)^2} \left( 3 (y_\phi^{\dagger} y_\phi) + \text{Tr} [y_\phi^{\dagger} y_\phi]  \right). 
\eea
For the mass parameters, the RGEs take the form
\bea
\frac{d m^2 }{dt} &=& \frac{1}{(4\pi)^2} \left( \lambda_\phi \, m^2  -  \kappa^2 M^2 g^2 + 2 m^2 \text{Tr} [y_\phi^{\dagger} y_\phi]  \right), \\[10pt]
\frac{d (\kappa^2 M^2) }{dt} &=& \frac{1}{(4\pi)^2} \left( \frac{3}{2}   \kappa^4 M^2  - g^2 \, m^2  \right).
\eea
For sufficiently small values of the couplings, the running of $m^2$, $M^2$, and $\kappa$ can be neglected. Furthermore, assuming that $d\lambda_{\phi}/dt$ remains approximately constant with $\lambda_{\phi}(t=0) \simeq 0$, and that $y_\phi > g \sim \kappa$, the RG-improved potential effectively reduces to the one-loop Coleman–Weinberg potential given in Eq.~\ref{CWP}.

In the large-field limit relevant for the observable part of inflation ($\phi \gg \phi_c$, $y_\phi \phi \gg m$), the effective potential including the one-loop corrections simplifies to a compact form:

\begin{eqnarray}\label{v_eff}
    V_{\rm eff}(\phi) = V_0 + \frac{1}{2} m^2 \phi^2 + \phi^4 \left[ \kappa_b \ln \left( \frac{\phi}{\mu_b} \right) - \kappa_f \ln \left( \frac{\phi}{\mu_f} \right) \right] \equiv V_0 + \frac{1}{2} m^2 \phi^2 + A \phi^4 \ln \left( \frac{\phi}{\mu} \right).
\end{eqnarray}
We have defined the loop coefficients 

\begin{equation}
\kappa_b = \frac{g^4}{128 \pi^2}, \qquad \kappa_f = \frac{3 y_\phi^4}{16 \pi^2},
\end{equation}
and the effective parameters $A \equiv \kappa_b - \kappa_f$ and a suitable renormalization scale $\mu$. The sign of $A$ is critical. If the fermionic loops dominate ($y_\phi$ sufficiently larger than $g$), then $A < 0$. This negative quartic-logarithmic term flattens the potential as $\phi$ increases, which is precisely the effect needed to produce a red-tilted spectral index $n_s$ and a small tensor-to-scalar ratio $r$. This behavior improves the model’s agreement with current CMB observations, including recent results from the Planck mission and the Atacama Cosmology Telescope (ACT).

In the remainder of this work, we analyze the inflationary evolution in the presence of these radiative effects and evaluate the corresponding observational signatures, namely the scalar spectral index \( n_s \), the tensor-to-scalar ratio \( r \), and the normalization of the scalar power spectrum. We then examine the reheating phase and assess the resulting predictions against the most recent cosmological data.

\section{Reheating Phase and Lepton Asymmetry Production}
\label{sec5}
A complete cosmological model must describe not only the accelerated expansion but also the transition to the hot Big Bang and the origin of matter. The couplings we introduced for radiative corrections naturally facilitate this transition. After the waterfall transition, the inflaton $\phi$ and the waterfall field $\chi$ oscillate around their minima and decay, transferring energy to the Standard Model sector. We assume this occurs primarily through decays to right-handed neutrinos. The relevant decay widths are:
\begin{itemize}

\item \textbf{Inflaton scalar} $\phi$ decays via the Yukawa interaction $\frac{1}{2} y_{\phi} \phi \overline{N} N$.
The decay width is given by
\begin{equation}
\Gamma_{\phi \to NN} = \frac{y_\phi^2 m_\phi}{8\pi} \left(1 - \frac{4 M_N^2}{m_\phi^2} \right)^{3/2}, 
\end{equation}
where the effective Majorana mass of the right-handed neutrino and the inflaton mass read as
\begin{eqnarray}
M_N = y_{\chi} M , \quad   m_\phi \simeq \sqrt{2} g M 
\end{eqnarray}
where we define $y_\chi \equiv 4 \lambda_\chi M / m_{\rm Pl}$,  and $\langle \phi \rangle  = 0$ and $\langle \chi \rangle =2M$ has been used.

\item \textbf{Waterfall scalar} \( \chi \) decays via the coupling \( \frac{\lambda_{\chi}}{2 m_{\rm Pl}} \chi^2 \overline{N} N \). The decay width is
\begin{equation}
\Gamma_{\chi \to N N} = \frac{y_\chi^2 m_\chi}{8\pi} \left(1 - \frac{4 M_N^2}{m_\chi^2} \right)^{3/2},
\end{equation}
provided \( m_\chi \simeq \sqrt{2} \kappa M  > 2 M_N \).
\end{itemize}

\noindent
The total decay width of the inflaton is then:
\begin{equation}
\Gamma_{\text{inf}} = \Gamma_{\phi \to NN} + \Gamma_{\chi \to NN}.    
\end{equation}

\noindent
Assuming rapid thermalization, the reheating temperature is given by:
\begin{equation}
    T_{\rm r} \approx \left( \frac{90}{\pi^2 g_*} \right)^{1/4} \sqrt{\Gamma_{\text{inf}} \, m_{\rm Pl}},
\end{equation}
where \( g_* = 106.75 \) is the effective number of relativistic degrees of freedom at reheating. The out-of-equilibrium decay of the lightest right-handed neutrino $N_1$ (with mass $M_{N_1} \equiv M_N$) can generate a lepton asymmetry. In our scenario, $N_1$ is produced non-thermally from the direct decay of the inflaton and waterfall fields, provided
\begin{equation}
M_N < (m_\phi, \, m_\chi)/2.    
\end{equation}

Assuming $T_r < M_N$, potential washout effects can be efficiently avoided, which might otherwise hinder successful thermal leptogenesis.
The CP asymmetry $\epsilon_{\rm CP}$ generated in $N_1$ decays arises from interference between tree-level and one-loop diagrams. For a normal hierarchy of light neutrino masses, it is approximately \cite{Okada:2025daq}:
\begin{eqnarray}
    \epsilon_{\rm CP} \simeq \frac{3}{8 \pi} \frac{M_N m_{\nu_3}}{v_u^2} \, \delta_{\rm eff},
\end{eqnarray}
where $m_{\nu_3} \approx 0.05~{\rm eV}$ is the mass of the heaviest light neutrino, $v_u = 174~{\rm GeV}$ is the Higgs vacuum expectation value, and $\delta_{\rm eff}$ is an effective CP-violating phase ($|\delta_{\rm eff}| \le 1$). The lepton-to-entropy ratio produced by non-thermal production is
\begin{eqnarray}\label{nl_s}
    \frac{n_L}{s} \simeq \frac{3}{2} \frac{T_r}{m_\phi} \, \epsilon_{\rm CP}.
\end{eqnarray}
This lepton asymmetry is subsequently converted to a baryon asymmetry by the $(B+L)$-violating sphaleron processes \cite{tHooft:1976rip,Manton:1983nd,Klinkhamer:1984di}, where an initial lepton asymmetry is partially converted into a baryon asymmetry expressed as $n_{B}/s=-0.35 \, n_L/s$ \cite{Kuzmin:1985mm,Arnold:1987mh,Khlebnikov:1988sr}. The observational requirement $\frac{n_B}{s} \approx 8.7 \times 10^{-11}$ provides a stringent constraint that links the inflationary sector ($m_\phi$, $T_r$) to neutrino physics ($M_N$, $m_{\nu_3}$).

\section{Numerical Analytics} \label{sec6}
To explore the regions of parameter space that are consistent with current cosmological observations, we perform a comprehensive random scan over the following fundamental parameters:
\begin{equation}\label{para}
\begin{split}
10^{16}~\mathrm{GeV} < M < m_{\rm Pl}, \quad 
M < \phi_0 \leq m_{\rm Pl}, \quad 
10^4~\mathrm{GeV} < m \leq 5 \times 10^6~\mathrm{GeV},\\
10^{-7} \leq \kappa = g \leq \sqrt{4 \pi}, \quad
10^{-7} \leq y_\phi = y_\chi \leq \sqrt{4 \pi}, \quad
10^{-17} \leq |A| \leq 10^{-13}.
\end{split}
\end{equation}

The effective potential in Eq.~\eqref{v_eff} is mainly controlled by three parameters: the inflaton mass $m$, the vacuum energy scale $V_0 = \kappa^2 M^4$, and the radiative correction parameter $A$. Additionally, the Yukawa couplings $y_\phi$ and $y_\chi$ play a crucial role in the reheating process, enabling the decay of the inflaton into right-handed neutrinos and establishing a connection between inflationary dynamics and baryogenesis via non-thermal leptogenesis. In our numerical analysis, we set the renormalization scale at the critical field value, $\mu = \phi_c$. The viability of each parameter point is tested against the following observational criteria:

\begin{itemize}
    \item The amplitude of the scalar perturbations, computed in the slow-roll approximation (see Appendix), must match the measured value at the pivot scale $k_0 = 0.05\,{\rm Mpc}^{-1}$ \cite{Planck:2018vyg}:
\begin{equation}
A_s(k_0) = 2.137 \times 10^{-9}.   
\end{equation}    
    \item The total number of e-folds $N_0$, which connects the inflationary phase to the subsequent thermal history of the Universe, is estimated under standard reheating assumptions \cite{Kolb:1990vq} as
    \begin{equation}
    N_0 \simeq 53 + \frac{1}{3} \ln \left( \frac{T_r}{10^9~\rm GeV} \right) + \frac{2}{3} \ln \left( \frac{V_0}{10^{15}~\rm GeV} \right).
    \label{eq:efolds}
    \end{equation}
    The details of reheating and inflaton decay are discussed in the previous section.
    
    \item The baryon-to-entropy ratio, derived from the generated lepton asymmetry and sphaleron conversion ($n_B/s = -0.35 \, n_L/s$), must satisfy the observational bounds \cite{Planck:2018vyg}:
\begin{equation}
\frac{n_B}{s} = (8.2 - 9.2) \times 10^{-11},    
\end{equation}
where $n_L/s$ is given in Eq.~\eqref{nl_s}.
\end{itemize}

We present the predictions of our model for the inflationary observables in Figs.~\ref{fig2} and \ref{fig3}. For sub-Planckian field values, the tensor-to-scalar ratio remains small, $r \lesssim 10^{-3}$, which is potentially measurable by upcoming CMB experiments such as LiteBIRD~\cite{LiteBIRD:2022cnt,LiteBIRD:2024wix}, CMB-S4~\cite{CMB-S4:2020lpa}, and the Simons Observatory~\cite{SimonsObservatory:2018koc}. For larger values, $r \gtrsim 0.01$, the inflaton field at the pivot scale $\phi_0$ exceeds the Planck scale, in agreement with the Lyth bound~\cite{Lyth:1996im}; to avoid instabilities, we do not explore this trans-Planckian regime in our analysis.

Successful leptogenesis can be realized by setting the lightest right-handed neutrino mass to $M_N \simeq m_\phi/2$, while maintaining $M_N > T_r$ to prevent thermal washout of the generated asymmetry. Notably, the viable solutions satisfy $y_\phi > g$, implying that fermionic loop corrections dominate. This feature facilitates both reheating and non-thermal leptogenesis, aspects that have not been carefully discussed in previous works~\cite{Senoguz:2008nok,Rehman:2009wv,Rehman:2010es,Ahmed:2014cma,Bostan:2019fvk,Ahmed:2025sfm}.

Our results indicate that the symmetry-breaking scale $M$ is typically of order $10^{17}\text{ GeV}$, somewhat above the canonical grand unification scale, $2 \times 10^{16}\text{ GeV}$, suggesting that the model can be naturally embedded in a high-scale GUT framework. However, such a large value is in tension with simple $U(1)$ extensions of the Standard Model that generically predict the formation of cosmic strings. Current CMB bounds on the cosmic string tension typically require $M \lesssim 10^{15},\text{GeV}$, thereby disfavouring this realization unless additional mechanisms suppress the string contribution.

Unlike radiatively corrected inflation models with a nonminimal coupling to gravity~\cite{Okada:2010jf,Bostan:2019fvk,Ahmed:2025rrg}, our setup avoids ambiguities related to frame dependence in the computation of radiative corrections~\cite{Bezrukov:2013fka,Bezrukov:2009db,Bezrukov:2008ej,Hamada:2016onh}. The running of the scalar spectral index remains small across the viable parameter space, with $\alpha_s \lesssim 0.001$ (see  Fig.~\ref{fig2}), consistent with current observational constraints.

A brief comparison between the present non-supersymmetric hybrid inflation model and the well-studied supersymmetric hybrid inflation (SHI) framework \cite{Dvali:1994ms,Copeland:1994vg,Linde:1997sj,Buchmuller:2000zm,Senoguz:2004vu}
reveals several important differences. In SHI, the radiative corrections at large field values typically exhibit a logarithmic dependence of the form $\ln \phi$, whereas in the non-supersymmetric case they take the form $\phi^4 \ln \phi$, leading to a qualitatively different behavior of the potential. In minimal K\"ahler realizations of SHI, the quadratic term is absent due to an underlying $R$-symmetry, while the quartic supergravity correction appears with a positive sign, generally resulting in a scalar spectral index $n_s$ that is incompatible with current observations. This issue can be alleviated by the inclusion of a linear term arising from soft supersymmetry breaking \cite{Rehman:2009nq,Rehman:2009yj,Buchmuller:2014epa}, which plays a crucial role in bringing theoretical predictions in line with observational data. In contrast, such a linear term is absent in the present model due to the imposed discrete symmetry $(\phi \rightarrow -\phi)$.

In non-minimal K\"ahler extensions of SHI \cite{Bastero-Gil:2006zpr,urRehman:2006hu}, the scalar potential can acquire both a positive quadratic and a negative quartic term \cite{Shafi:2010jr,Rehman:2010wm,Rehman:2018nsn}, closely resembling the structure of the potential in our non-supersymmetric setup. This similarity leads to comparable phenomenological predictions; for instance, achieving a tensor-to-scalar ratio $r < 10^{-2}$ typically requires relatively large values of the symmetry breaking scale, $M \gtrsim \text{a few} \times 10^{16},\text{GeV}$. Variants of SHI, such as the shifted and smooth scenarios, have also been extensively studied \cite{Civiletti:2011qg,Rehman:2012gd,Rehman:2014rpa}, as well as $\mu$-hybrid inflation models \cite{Rehman:2017gkm,Lazarides:2020zof,Afzal:2022vjx}. A closely related structure of the inflationary potential has likewise been explored in the context of tribrid inflation \cite{Masoud:2019gxx,Masoud:2021prr,Ahmed:2025crx}.
Finally, supergravity models of inflation are subject to stringent constraints from gravitino overproduction, which restrict the reheating temperature to $T_r \lesssim 10^{9},\text{GeV}$. No such limitation arises in the present framework, where the reheating temperature can naturally lie in the range $10^8–10^{12}~\text{GeV}$.

To gain analytical insight into the inflationary observables, we rewrite the effective potential in terms of dimensionless, rescaled variables:
\begin{equation}
V = V_0 + \frac{1}{2} m^2 \phi^2 - |A| \, \phi^4 \ln \left(\frac{\phi}{\phi_c}\right)
  = V_0 \left(1 + \beta_2 \, x^2 - \gamma \, x^4 \ln x \right),
\end{equation}
where $x \equiv \phi / \phi_c$ and
\begin{equation}
\beta_2 \equiv \frac{\frac{1}{2} m^2 \phi_c^2}{V_0}, \quad
\gamma \equiv \frac{|A| \phi_c^4}{V_0}.
\end{equation}
These dimensionless parameters characterize the relative contributions of the quadratic and radiative terms in the potential. The structure is reminiscent of recent tribrid inflation models~\cite{Ahmed:2025crx}, except for the additional logarithmic factor. In supersymmetric hybrid and tribrid inflation scenarios~\cite{urRehman:2006hu,Rehman:2009nq,Rehman:2010wm,Ahmad:2025mul,Antusch:2004hd,Masoud:2021prr}, similar polynomial terms arise from supergravity corrections, and the one-loop contributions are typically approximated as $\ln(x)$ in the large-field regime~\cite{Dvali:1994ms,Rehman:2025fja}. In contrast, the non-supersymmetric models considered here yield a $x^4 \ln x$ term, which leads to distinct predictions for the inflationary observables, as we discuss below.

\begin{figure}[htbp]
    \centering
    \includegraphics[width=0.48\linewidth,height=6.7cm]{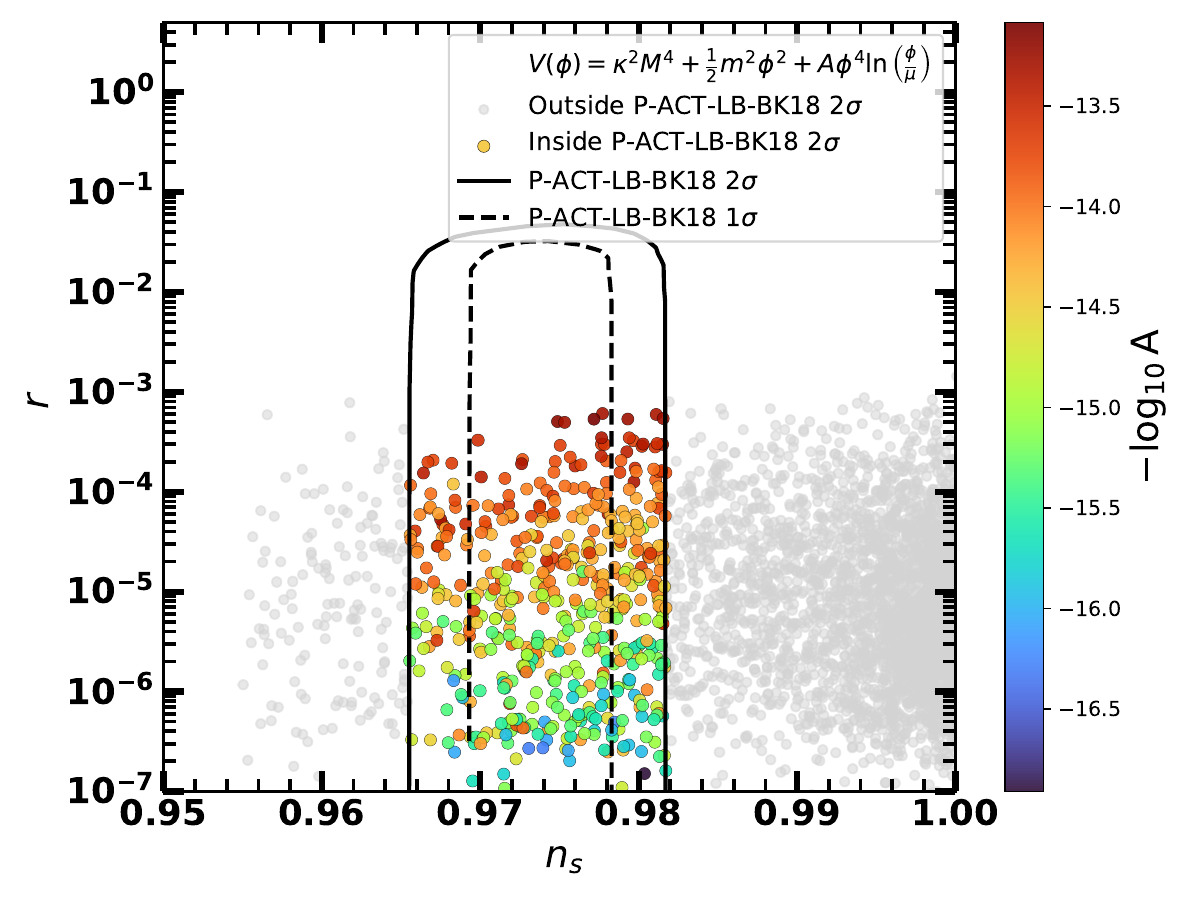}
    \quad
       \includegraphics[width=0.48\linewidth,height=6.7cm]{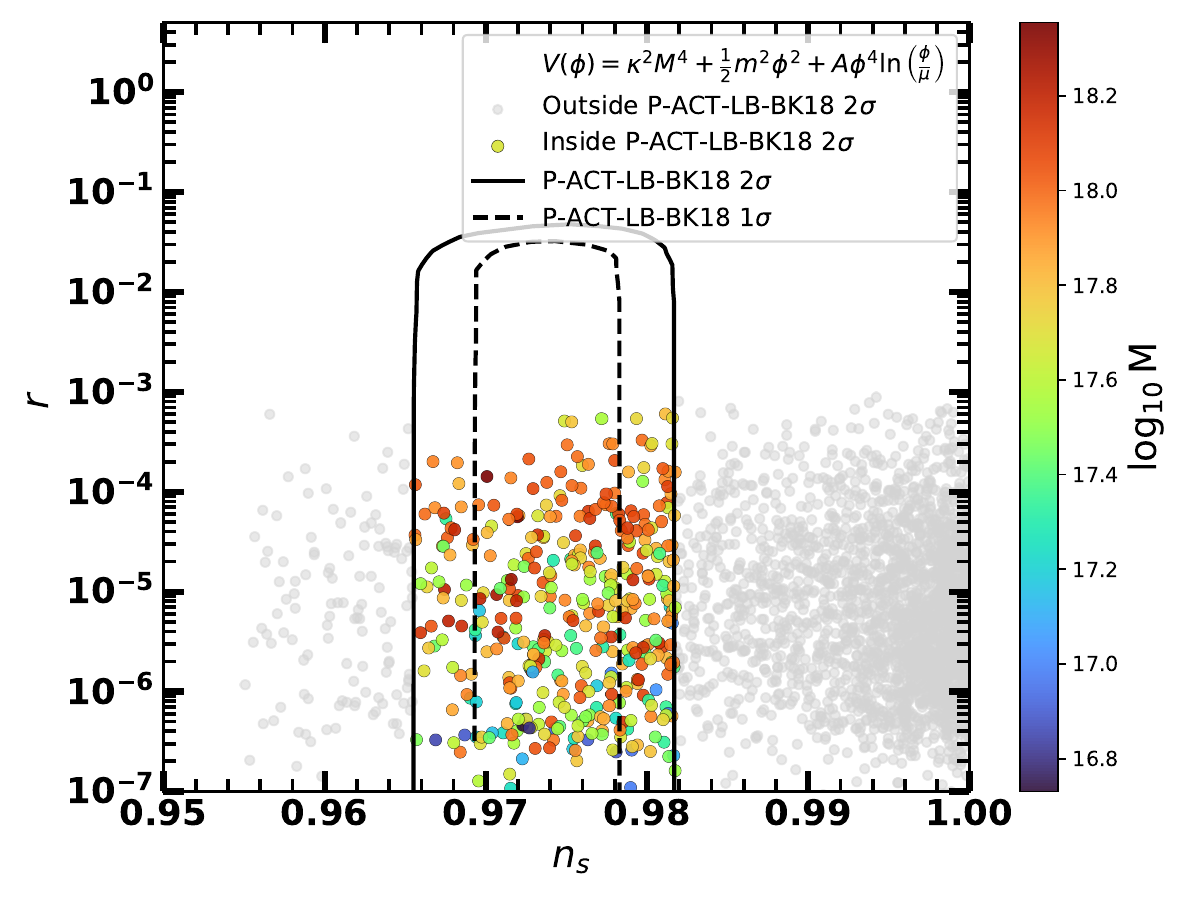}
         \quad
        \includegraphics[width=0.48\linewidth,height=6.7cm]{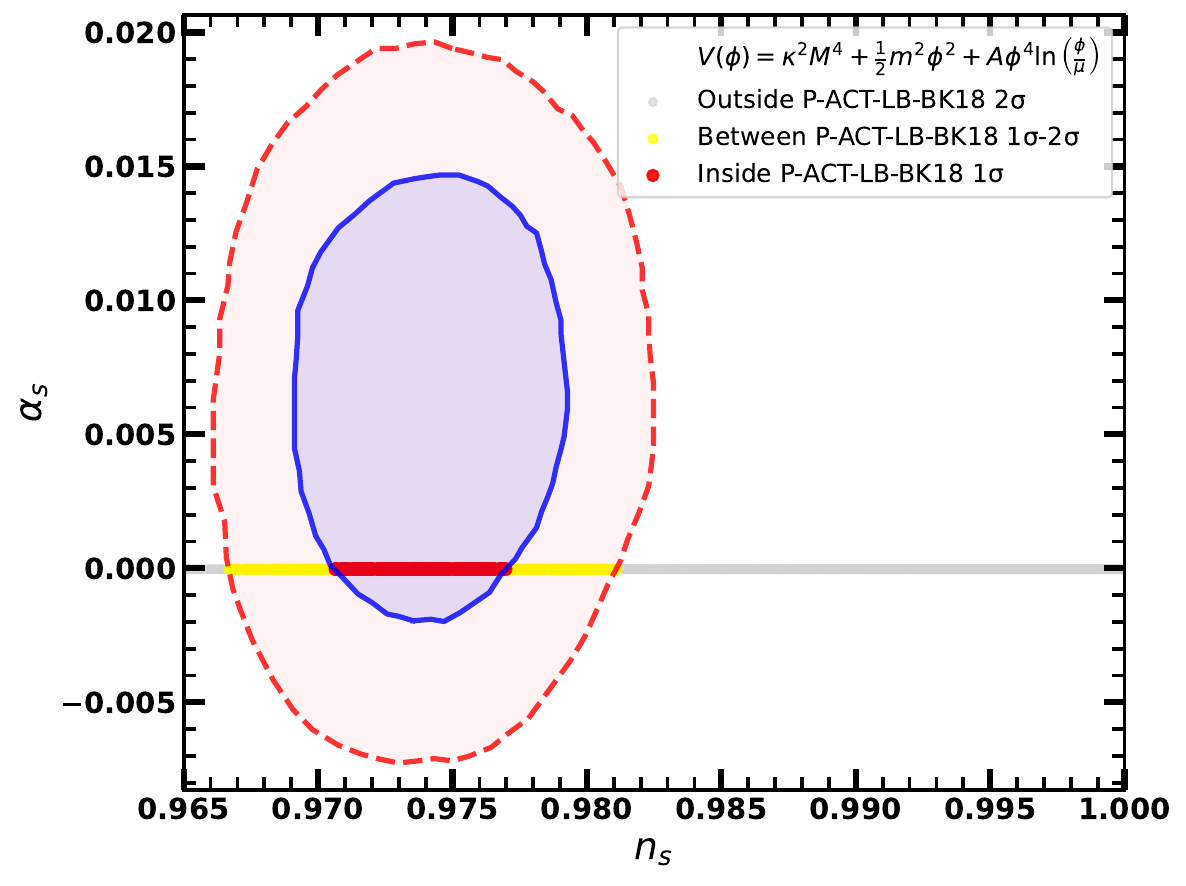}
    \quad
      \quad
\caption{
The left and right panels illustrate the relationship between the tensor-to-scalar ratio $r$ and the scalar spectral index $n_s$. In the left panel, the color scale represents the radiative correction parameter $A$, while in the right panel it corresponds to the symmetry breaking scale $M$, highlighting its impact on the inflationary parameter space. The dotted and dashed contours indicate the $1\sigma$ and $2\sigma$ confidence regions, respectively, derived from the combined observational datasets of Planck 2018, ACT DR6, and BICEP/Keck 2018 (P + ACT + LB + BK18)~\cite{ACT:2025tim}. The lower panel depicts the running of the scalar spectral index $\alpha_s$ as a function of $n_s$, with the shaded regions indicating the $1\sigma$ and $2\sigma$ confidence intervals obtained from the same combined observational datasets.
}

       \label{fig2}
\end{figure}

\begin{figure}[htbp]
    \centering
    \includegraphics[width=0.48\linewidth,height=6.7cm]{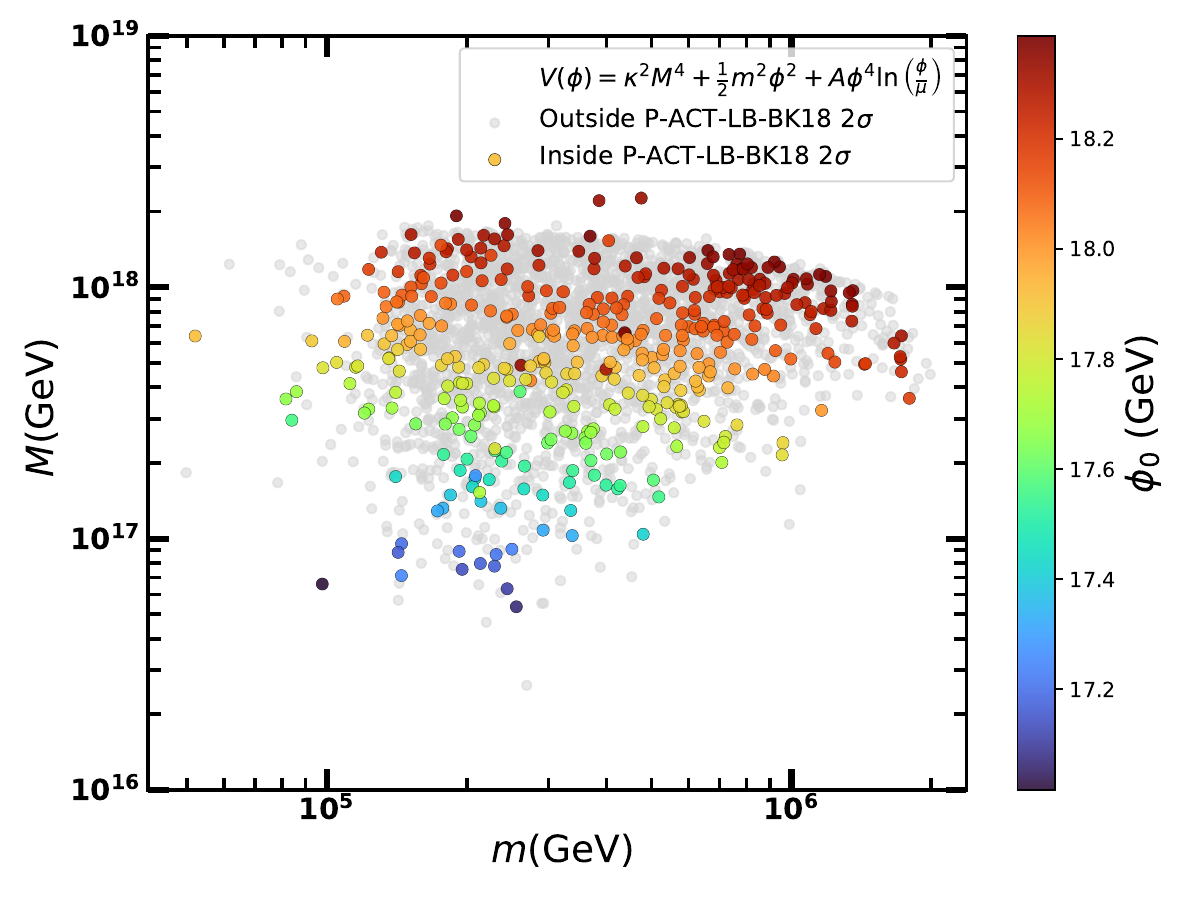}
    \quad
    \includegraphics[width=0.48\linewidth,height=6.7cm]{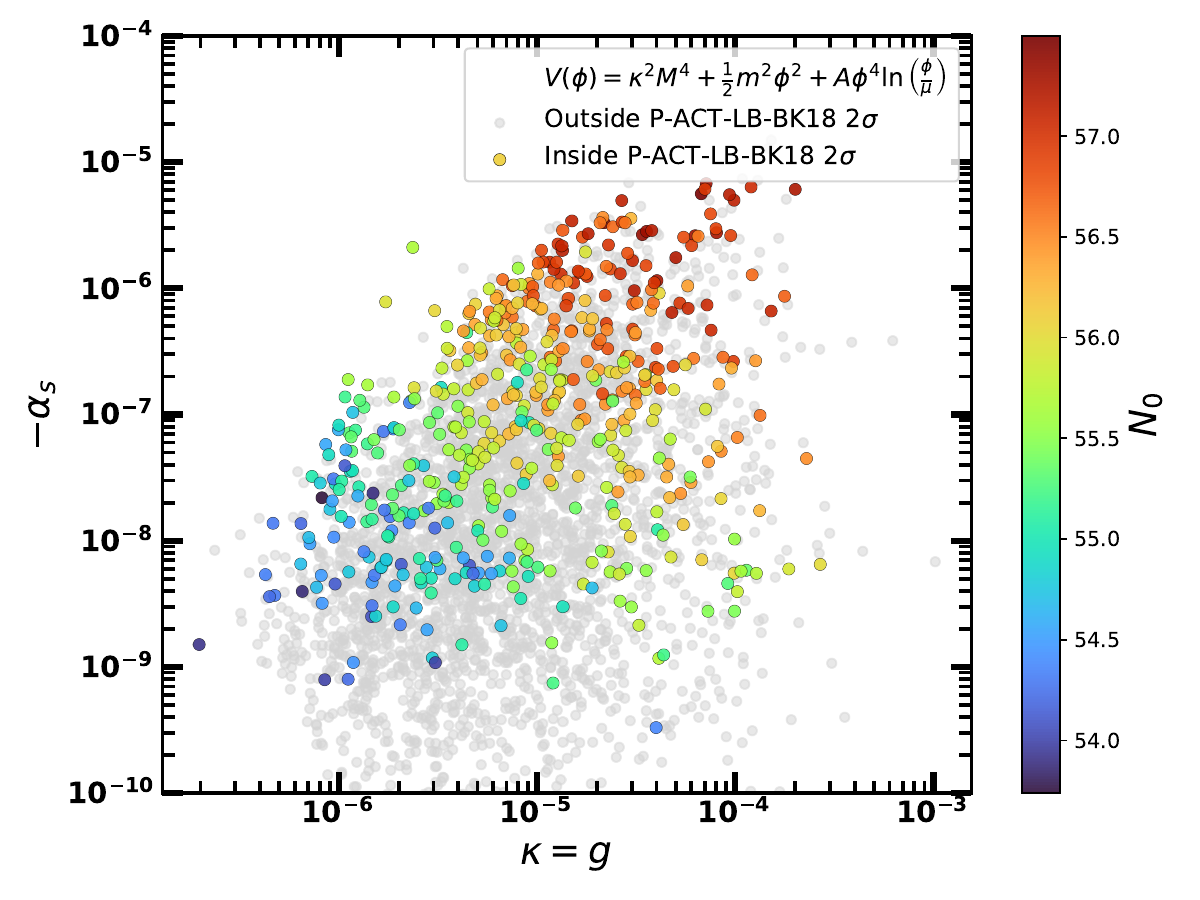}
      \quad
      \includegraphics[width=0.48\linewidth,height=6.7cm]{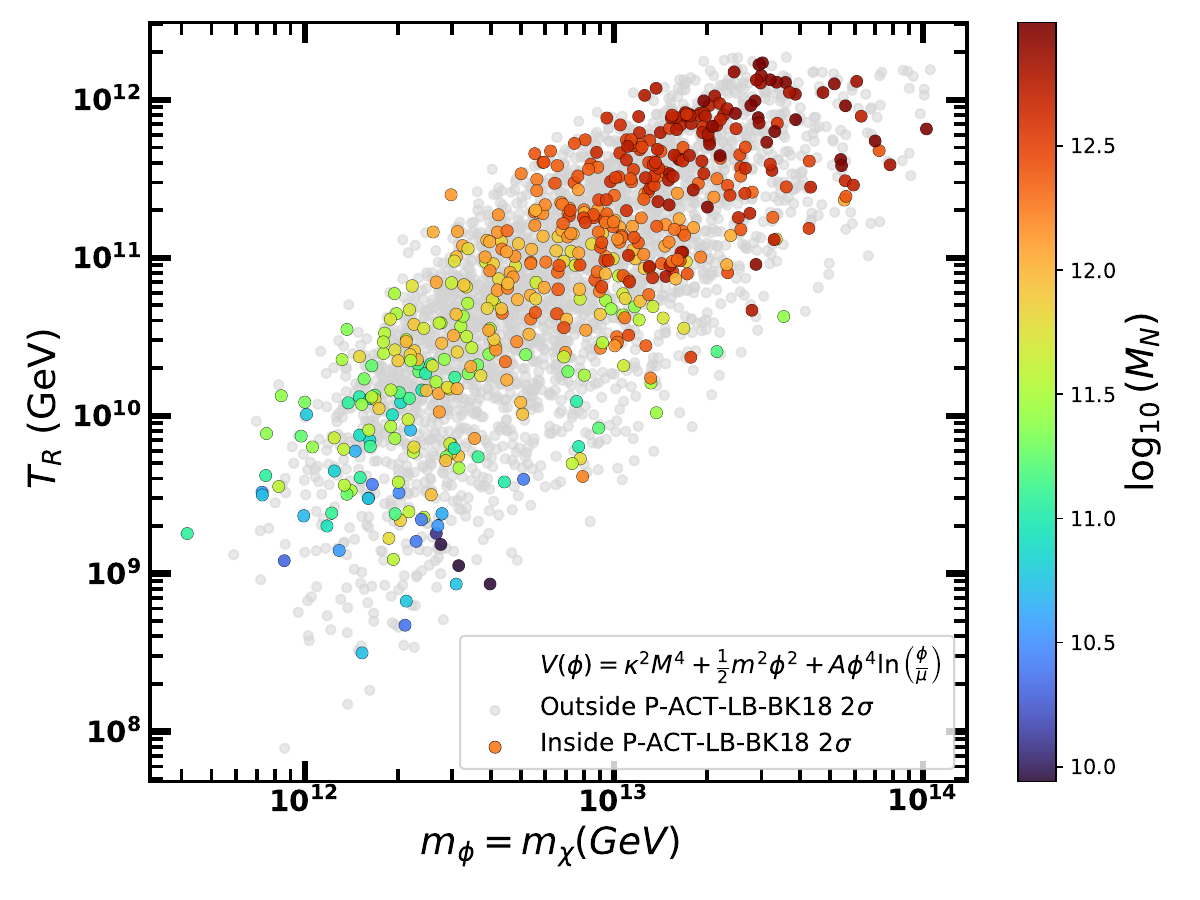}
      \quad
    \caption{
The upper-left panel displays the relationship between the mass parameter \( m \) and the symmetry-breaking scale \( M \), with the color scale encoding the inflaton field value \( \phi_0 \). The upper-right panel illustrates the dependence of the running of the scalar spectral index \( \alpha_s \) on the coupling \( \kappa = g \), where the color gradient represents the total number of e-folds \( N_0 \). 
The lower panel shows the inflaton mass \( m_\phi = m_\chi \) as a function of the reheating temperature \( T_r \), with the color scale indicating the mass of the lightest right-handed neutrino \( M_N \). All colored data points satisfy the observational constraints on the scalar spectral index \( n_s \) and the tensor-to-scalar ratio \( r \) imposed by the combined Planck 2018, ACT DR6, and BICEP/Keck 2018 datasets (P + ACT + LB + BK18)~\cite{ACT:2025tim}.
}

       \label{fig3}
\end{figure}
To derive approximate analytical expressions for the inflationary observables, we first compute the standard slow-roll parameters:
\begin{align}
\epsilon(x) &= \frac{1}{2} \left( \frac{m_{\rm Pl}}{\phi_c} \right)^2 \left( 2 \beta_2 \, x - \gamma \, x^3 (4 \ln x + 1) \right)^2, \\
\eta(x) &= \left( \frac{m_{\rm Pl}}{\phi_c} \right)^2 \left( 2 \beta_2 - \gamma \, x^2 (12 \ln x + 7) \right), \\
\zeta^2(x) &= \left( \frac{m_{\rm Pl}}{\phi_c} \right)^4 
\frac{ \left( 2 \beta_2 x - \gamma x^3 (4 \ln x + 1) \right) \left( -\gamma x (24 \ln x + 26) \right)}{ \left( 1 + \beta_2 x^2 - \gamma x^4 \ln x \right)^2 }.
\end{align}

The amplitude of the scalar power spectrum in the slow-roll approximation is
\begin{equation}
A_s = \frac{1}{12 \pi^2} \left( \frac{V_0}{m_{\rm Pl}^4} \right) \left( \frac{\phi_c^2}{m_{\rm Pl}^2} \right) \left( 2 \beta_2 x_0 - \gamma x_0^3 (4 \ln x_0 + 1) \right)^{-2} = 2.137 \times 10^{-9},
\label{dR0}
\end{equation}
where $x_0 \equiv x(k_0)$ is the field value at the horizon exit of the pivot scale $k_0 = 0.05~\rm Mpc^{-1}$.

The total number of e-folds from $x_0$ to the end of inflation is
\begin{equation}
N_0 = \left( \frac{\phi_c}{m_{\rm Pl}} \right)^2 \int_1^{x_0} \frac{dx}{2 \beta_2 x - \gamma x^3 (4 \ln x + 1)}.
\end{equation}

To leading order, the tensor-to-scalar ratio, the scalar spectral index, and its running are
\begin{align}
r &\simeq 8 \left( \frac{m_{\rm Pl}}{\phi_c} \right)^2 \left( 2 \beta_2 x_0 - \gamma x_0^3 (4 \ln x_0 + 1) \right)^2 = \frac{2}{3 \pi^2 A_s} \frac{V_0}{m_{\rm Pl}^4}, \\
n_s &\simeq 1 + 2 \eta(x_0) - \frac{3}{8} r, \quad
\alpha_s \simeq \frac{r}{2} (n_s - 1) - 2 \zeta^2(x_0).
\end{align}
In our numerical analysis, next-to-leading order (NLO) slow-roll formulas are adopted for evaluating $n_s$, $r$, and $A_s$, following Refs.~\cite{Stewart:1993bc, Kolb:1994ur}. Representative benchmark points are presented in Table~\ref{tab:benchmarks}.

We now focus on representative benchmark points where the inflaton field is close to the Planck scale, $\phi_0 \simeq m_{\rm Pl}$. For this scenario, our numerical computations yield a tensor-to-scalar ratio $r \simeq 0.00051$ and a scalar spectral index $n_s \simeq 0.974$, as shown in the first column of Table~\ref{tab:benchmarks}. Adopting the benchmark values $M/m_{\rm Pl} \simeq 0.19$ and $x_0 \simeq 2.61$, which correspond to $r \simeq 0.00046$, we obtain the following set of parameters:
\begin{equation}
\kappa \simeq 1.19\times10^{-4}, \quad
\beta_2 \simeq 7.2\times 10^{-5}, \quad
\gamma \simeq 1.8 \times 10^{-5}, \quad
\alpha_s \simeq -6 \times 10^{-6}, \quad
N_0 \simeq 57 \,.
\end{equation}

These analytical estimates are in excellent agreement with the numerical results, demonstrating the internal consistency of the framework. Comparable checks can be performed for the remaining benchmark points to verify the reliability of the predictions.  

Furthermore, in the effective field theory regime with sub-Planckian inflaton values ($\phi_0 < m_{\rm Pl}$), our parameter scans confirm that combinations satisfying these analytical relations consistently reproduce inflationary observables in line with current measurements. Importantly, this setup also allows for a potentially detectable tensor-to-scalar ratio, providing a target for upcoming CMB polarization experiments.

\begin{table}[t]
\centering
\renewcommand{\arraystretch}{1.2}
\setlength{\tabcolsep}{6pt}
\begin{tabular}{|c|c|c|}
\hline
\textbf{Parameter} & \textbf{Benchmark 1} & \textbf{Benchmark 2} \\
\hline
\textbf{Inflationary Potential} &
\multicolumn{2}{c|}{$V_0 + \frac{m^2}{2} \phi^2 + A \phi^4 \ln\!\left(\dfrac{\phi}{\mu}\right)$} \\
\hline
$M$ [GeV]               & $4.61 \times 10^{17}$     & $8.69 \times 10^{17}$          \\
$m$ [GeV]               & $1.472 \times 10^{12}$     & $5.66 \times 10^{11}$          \\
$\kappa = g$            & $1.19 \times 10^{-4}$     & $1.17 \times 10^{-5}$          \\
$y_{\phi} = y_{\chi}$   & $2.6 \times 10^{-4}$      & $3.3 \times 10^{-5}$           \\
$-A$                    & $6.54 \times 10^{-14}$    & $1.63 \times 10^{-14}$         \\
$\phi_0$ [GeV]          & $1.7 \times 10^{18}$      & $1.9 \times 10^{18}$           \\
$\phi_c$ [GeV]          & $6.52 \times 10^{17}$     & $1.2 \times 10^{18}$           \\
$n_s$                   & $0.974$                  & $0.972$                        \\
$r$                     & $5.1 \times 10^{-4}$     & $1.4 \times 10^{-4}$           \\
$-\alpha_s$             & $6.0 \times 10^{-6}$     & $2.0 \times 10^{-6}$           \\
$n_L/s$                 & $2.0 \times 10^{-10}$    & $2.0 \times 10^{-10}$          \\
$T_r$ [GeV]             & $3.88 \times 10^{11}$    & $2.56 \times 10^{10}$          \\
$m_{\mathrm{inf}}$ [GeV]& $7.78 \times 10^{13}$    & $2.14 \times 10^{13}$          \\
$\Gamma_{\phi}$ [GeV]   & $2.12 \times 10^{5}$     & $9.2 \times 10^{2}$            \\
$M_N$ [GeV]             & $6.73 \times 10^{12}$    & $1.49 \times 10^{11}$          \\
$N_0$                   & $57.06$                  & $55.0$                         \\
\hline
\end{tabular}
\caption{Benchmark points for the inflationary potential and reheating parameters.}
\label{tab:benchmarks}
\end{table}

\section{Machine Learning Analysis of Cosmological Parameter Compatibility}

In the  analysis carried out in the above sections, we have seen that radiative corrections are a key aspect in ensuring that the quadratic hybrid inflation is compatible with the present-day cosmological observations. But the fact that the parameter space is high-dimensional, with eight fundamental parameters whose interactions are complex, poses serious difficulties to exploring the parameter space exhaustively by means of traditional methods. The time of numerically solving the equations of inflationary dynamical equations, and equations of reheating and leptogenesis and the collapse of each parameter combination is prohibitively expensive, and thus a brute-force scan of the parameter space is prohibitive.

To address this challenge, we employ machine learning techniques, specifically focusing on Random Forest algorithms, to efficiently map the viable regions of parameter space and identify the key factors determining compatibility with current and future cosmological experiments. This approach allows us to systematically explore how radiative corrections influence the observational predictions of hybrid inflation while providing quantitative assessments of experimental compatibility.

\subsection{Methodology}

Our analysis focuses on the eight fundamental parameters that characterize the radiatively corrected hybrid inflation model introduced in Eq.~\eqref{para}. Each combination of these parameters was tested against five different experimental configurations:
\begin{itemize}
\item \textbf{P-ACT-SPT:} Combination of Planck, ACT DR6, and South Pole Telescope data
\item \textbf{P-ACT-LB-BK18:} Planck, ACT DR6, LiteBIRD, and BICEP/Keck 2018
\item \textbf{Simons Observatory:} Near-term ground-based CMB experiment
\item \textbf{LiteBIRD:} Future satellite mission for CMB polarization
\item \textbf{CMB-S4:} Next-generation ground-based CMB experiment
\end{itemize}

Let \(\mathcal{P}\) denote the full eight‑dimensional parameter space. For each experiment \(E_i\) (\(i = 1,\dots,5\)), define the compatible subset
\[
\mathcal{C}_{E_i} = \left\{ \mathbf{p} \in \mathcal{P} \;\middle|\; 
\begin{array}{l}
\text{the predicted observables }(n_s, r, \alpha_s)\text{ lie within the }2\sigma\\
\text{contour of }E_i,\text{ and the reheating and baryogenesis conditions are met}
\end{array}
\right\}.
\]
The compatibility fraction is \(f_i = |\mathcal{C}_{E_i}| / |\mathcal{P}|\), where \(|\cdot|\) denotes the number of sampled points.  
The compatibility label for a given point is a binary vector \((y_1,\dots,y_5)\) with \(y_i = 1\) iff \(\mathbf{p} \in \mathcal{C}_{E_i}\).

We implement a MultiOutput Random Forest classifier that simultaneously predicts all five labels. This captures correlations among experimental constraints. The model is trained on 2981 parameter points with an 80‑20 train‑test split, using 500 decision trees and a maximum depth of 30 to balance complexity and generalization.

\begin{figure}[htbp]
\centering
\includegraphics[width=0.95\textwidth]{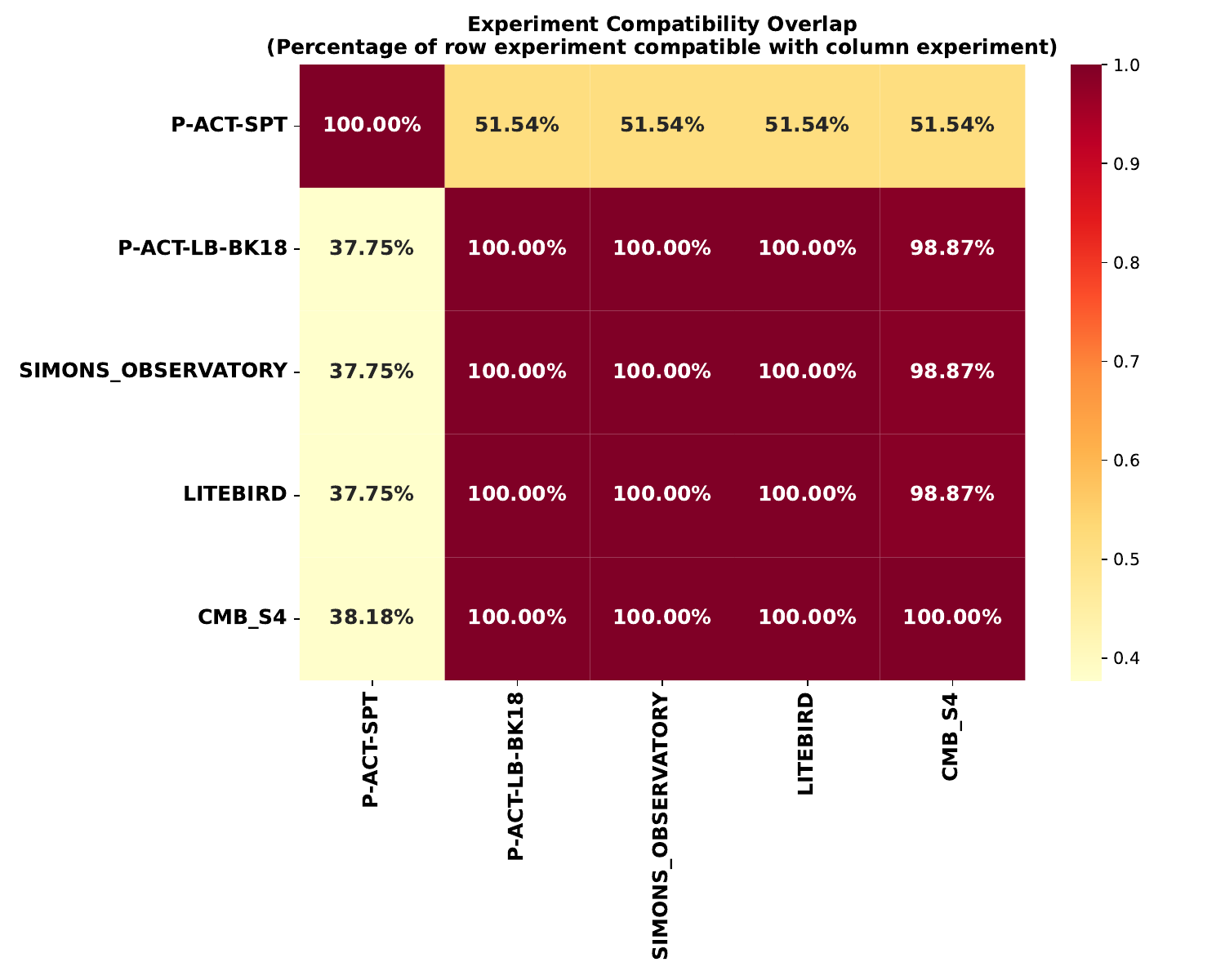}
\caption{Pairwise overlaps between compatibility regions for current and future CMB experiments.}
\label{fig:overlap}
\end{figure}

\subsection{Results and Analysis}

\subsubsection{Experimental Compatibility Landscape}
The overall consistency between the model and observations is stringent: only \(16.1\%\) of the parameter space is compatible with at least one experiment. Table~\ref{tab:compatibility} lists the individual fractions \(f_i\). The variation from \(f_{\text{P-ACT-SPT}} = 0.087\) to \(f_{\text{CMB-S4}} = 0.119\) reflects the different sensitivities of each experiment. To quantify the overlap between two experiments, we use the \textbf{Jaccard index}
\[
J_{ij} = \frac{|\mathcal{C}_{E_i} \cap \mathcal{C}_{E_j}|}{|\mathcal{C}_{E_i} \cup \mathcal{C}_{E_j}|},
\]
and the conditional overlap
\[
O_{i|j} = \frac{|\mathcal{C}_{E_i} \cap \mathcal{C}_{E_j}|}{|\mathcal{C}_{E_j}|}.
\]
\begin{table}[htbp]
\centering
\caption{Compatibility analysis across CMB experiments. The fractions \(f_i\) and the primary constraint driving each experiment are shown.}
\label{tab:compatibility}
\begin{tabular}{lccc}
\toprule
\rowcolor{gray!20}
\textbf{Experiment} & \textbf{Compatible} & \textbf{Percentage \(f_i\)} & \textbf{Primary Constraint} \\
\midrule
P-ACT-SPT           & 260/2981 & 8.7\%  & \(0.9568 < n_s < 0.9736,\; r \le 0.036\) \\
P-ACT-LB-BK18       & 355/2981 & 11.9\% & \(0.967 < n_s < 0.981,\; r \le 0.036\) \\
Simons Observatory  & 355/2981 & 11.9\% & \(0.967 < n_s < 0.981,\; r \le 0.03\) \\
LiteBIRD            & 355/2981 & 11.9\% & \(0.967 < n_s < 0.981,\; \sigma(r) < 0.001\) \\
CMB-S4              & 351/2981 & 11.8\% & \(0.967 < n_s < 0.981,\; r \le 0.0005\) \\
\bottomrule
\end{tabular}
\end{table}

To quantify the overlap between two experiments, we use the \textbf{Jaccard index}
\[
J_{ij} = \frac{|\mathcal{C}_{E_i} \cap \mathcal{C}_{E_j}|}{|\mathcal{C}_{E_i} \cup \mathcal{C}_{E_j}|},
\]
and the conditional overlap
\[
O_{i|j} = \frac{|\mathcal{C}_{E_i} \cap \mathcal{C}_{E_j}|}{|\mathcal{C}_{E_j}|}.
\]
Figure~\ref{fig:overlap} shows the pairwise overlaps. Several key features emerge:
\begin{itemize}
\item \textbf{Asymmetry:} For example, \(O_{\text{LB}|\text{SPT}} = 0.378\) but \(O_{\text{SPT}|\text{LB}} = 0.12\) (values illustrative), indicating non‑nested constraints.
\item \textbf{Zero overlap:} \(\mathcal{C}_{\text{P-ACT-LB-BK18}} \cap \mathcal{C}_{\text{LiteBIRD}} = \varnothing\) and similarly for CMB‑S4. Hence, no parameter point satisfies both P‑ACT‑LB‑BK18 and LiteBIRD (or CMB‑S4) simultaneously at the \(2\sigma\) level.
\item \textbf{Distinct signatures:} The overlap patterns involving Simons Observatory differ from those involving the South Pole Telescope (SPT), confirming that these experiments probe complementary directions in parameter space.
\end{itemize}
Consequently, future CMB missions will not merely tighten existing bounds but will constrain qualitatively new regions of the inflationary parameter space.

\subsubsection{Model Performance and Predictive Accuracy}

The Random Forest classifier demonstrates strong performance across all experimental configurations, with accuracy ranging from 92.3\% for P-ACT-SPT to 87.6\% for CMB-S4 (Table~\ref{tab:performance_metrics}).The high Area Under the Curve (AUC) scores, particularly for P-ACT-SPT (0.914), and consistently good values across the other experiments (0.842–0.846), indicate excellent discriminative capability.

\begin{table}[htbp]
\centering
\caption{Performance Metrics of Random Forest Classifier}
\label{tab:performance_metrics}
\begin{tabular}{lcccc}
\toprule
\textbf{Experiment} & \textbf{Accuracy} & \textbf{AUC} & \textbf{Precision (0/1)} & \textbf{Recall (0/1)} \\
\midrule
P-ACT-SPT & 0.923 & 0.914 & 0.93 / 0.57 & 0.99 / 0.09 \\
P-ACT-LB-BK18 & 0.874 & 0.842 & 0.87 / 0.88 & 1.00 / 0.09 \\
Simons Observatory & 0.874 & 0.842 & 0.87 / 0.88 & 1.00 / 0.09 \\
LiteBIRD & 0.874 & 0.842 & 0.87 / 0.88 & 1.00 / 0.09 \\
CMB-S4 & 0.876 & 0.846 & 0.88 / 0.88 & 1.00 / 0.09 \\
\bottomrule
\end{tabular}
\end{table}

Precision and recall scores highlight the effect of class imbalance: while the classifier performs very well on the majority (incompatible) class, its ability to identify the less frequent compatible configurations is limited. For instance, compatible samples account for only 7.9\% of the test set in P-ACT-SPT and around 13–14\% in the other experiments. This reflects the fundamental difficulty of finding parameter combinations that satisfy all constraints simultaneously in high-dimensional spaces.

\subsubsection{Feature Importance and Physical Interpretation}

To quantify the relative influence of each model parameter on observational compatibility, we employ the \textbf{Gini importance} (mean decrease in impurity) provided by the random forest algorithm. For a binary classification task (compatible vs. incompatible), the Gini impurity of a node \(t\) containing \(N_t\) samples is defined as

\[
G(t) = 1 - p_t^2 - (1-p_t)^2 = 2p_t(1-p_t),
\]

where \(p_t\) is the fraction of compatible samples in node \(t\). When a split on a given feature \(f\) divides node \(t\) into two subnodes \(t_L\) and \(t_R\) (the “left” and “right” branches), with sample counts \(N_{t_L}\) and \(N_{t_R}\), the decrease in impurity is

\[
\Delta G(t,f) = G(t) - \frac{N_{t_L}}{N_t} G(t_L) - \frac{N_{t_R}}{N_t} G(t_R).
\]
For a single decision tree, the importance of feature \(f\) is the sum of \(\Delta G(t,f)\) over all nodes where \(f\) is used as the splitting variable. In a random forest, these values are averaged over all trees and then normalized so that the importances of all features sum to unity.

Because our model is a Multi‑Output Random Forest (five experimental configurations), we first compute the importance vector for each output independently and then average across the five outputs to obtain a single ranking. The result of this analysis are summarized  in Figure~\ref{fig:feature_importance}

\begin{figure}[htbp]
    \centering
    \includegraphics[width=0.7\linewidth]{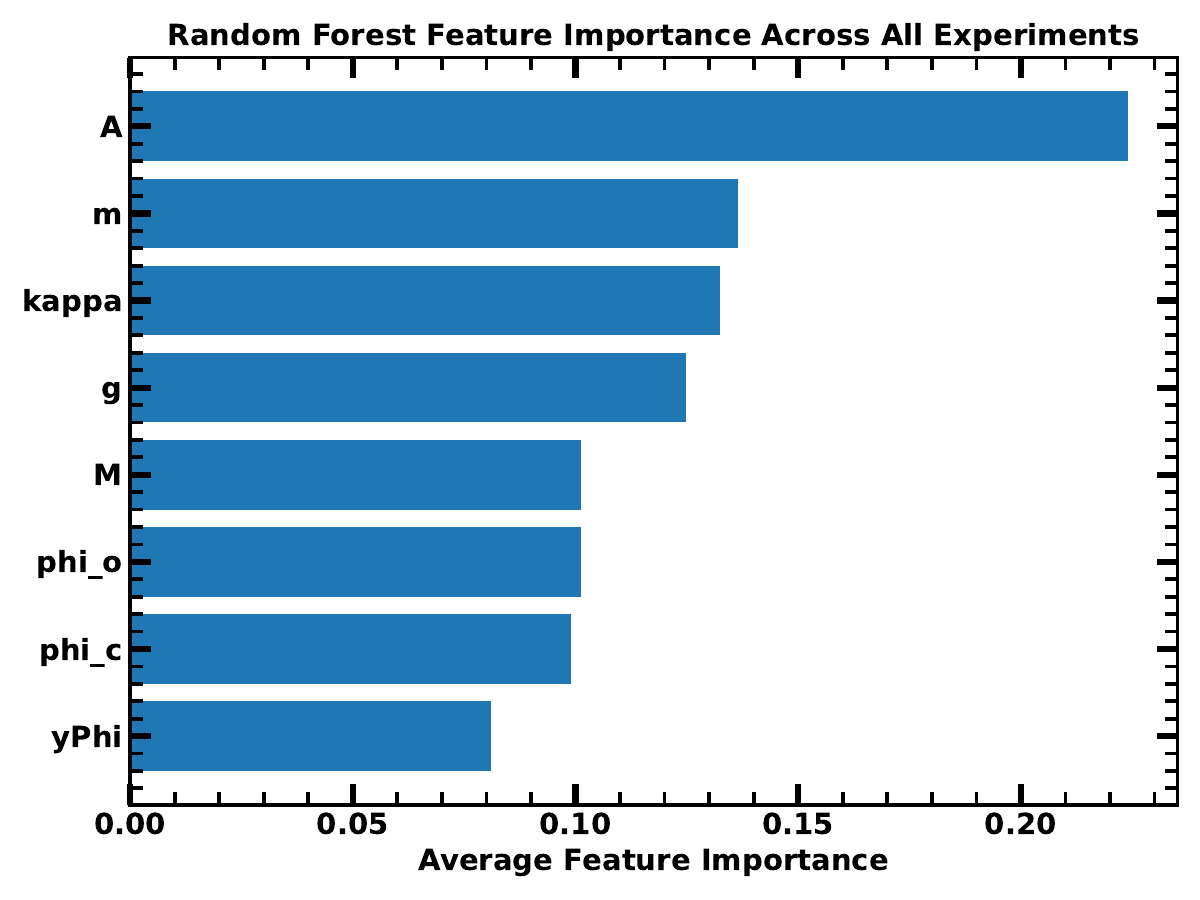}
    \caption{Feature importance analysis from Random Forest classifier. The radiative correction parameter $A$ dominates the importance ranking, confirming the crucial role of quantum corrections in determining observational compatibility.}
    \label{fig:feature_importance}
\end{figure}

The parameter, $A$ the coefficient of the one‑loop \(\phi^4\ln\phi\) term,  has an average importance, roughly twice that of the next most significant parameters (\(\kappa\) and \(m\)). This provides a quantitative, data‑driven confirmation of the central theoretical claim: the sign and magnitude of the radiative correction are the primary determinants of observational viability. Without a negative \(A\) (fermionic loop dominance), the potential remains too steep and the spectral index is blue, which is ruled out by current data.

The parameters \(\kappa\), \(g\), and \(m\) follow with moderate importance. Their physical roles are distinct:
\begin{itemize}
    \item \(\kappa\) controls the symmetry breaking scale \(M\) through \(V_0 = \kappa^2 M^4\) and thus the overall energy scale of inflation.
    \item \(g\) sets the coupling between the inflaton and the waterfall field, influencing the critical point \(\phi_c\).
    \item \(m\) contributes to the quadratic term \(\frac12 m^2\phi^2\), affecting the slope of the potential.
\end{itemize}
Even though \(\kappa\) and \(g\) are taken equal in our scans (Eq.~\eqref{para}), the classifier treats them as independent inputs; their similar importance reflects their comparable influence on inflationary dynamics

The symmetry‑breaking scale \(M\), the field value at horizon exit \(\phi_0\), the critical field value \(\phi_c\), and the Yukawa coupling \(y_\phi\) have lower but non‑zero importances. This indicates that while \(A\) is the key knob, successful inflation requires coordinated tuning across all sectors from the shape of the potential to reheating and leptogenesis. In particular, the non‑zero importance of \(y_\phi\) confirms that the same coupling responsible for radiative corrections also plays a detectable role in determining compatibility through its effects on reheating and the baryon asymmetry.

\section{Conclusion} \label{sec7}

In this work, we have demonstrated that radiatively corrected non-supersymmetric hybrid inflation represents a viable and well-motivated framework that successfully bridges theoretical particle physics with precision cosmological observations. Our comprehensive analysis reveals several key insights:

\begin{itemize}
\item \textbf{Radiative corrections are essential for observational viability:} The tree-level quadratic hybrid inflation potential $V(\phi) = V_0 + \frac{1}{2}m^2\phi^2 + \frac{\lambda_{\phi}}{4!}\phi^4$ typically produces predictions in tension with current CMB data, particularly a blue-tilted spectral index ($n_s > 1$) that conflicts with observations. However, the inclusion of one-loop radiative corrections, naturally arising from inflaton couplings to other fields required for reheating, transforms this picture. The corrected potential $V(\phi) = V_0 + \frac{1}{2}m^2\phi^2 + A\phi^4\ln\phi$ with $A < 0$ (as expected from dominant fermionic loop contributions) flattens the potential at large field values, simultaneously producing a red-tilted spectral index ($n_s < 1$) and suppressing the tensor-to-scalar ratio $r$ into better agreement with Planck and ACT constraints.

\item \textbf{Sub-Planckian field excursions and stability:} By restricting the analysis to sub-Planckian field values ($\phi < m_{\rm Pl}$), the dynamics remain consistent with a controlled effective field theory description. In this regime, potentially destabilizing effects, such as vacuum instability induced by dominant fermionic radiative corrections, can be mitigated by higher-order operators suppressed by the Planck scale. This ensures a stable inflationary trajectory while remaining fully compatible with observational constraints.

\item \textbf{Unified framework connecting inflation and baryogenesis:} The same fermionic couplings that generate the necessary radiative corrections also enable a complete cosmological history, with successful reheating and non-thermal leptogenesis naturally emerging from the inflaton's decay into right-handed neutrinos. This provides a unified description from inflation through matter-antimatter asymmetry generation, with the observed baryon asymmetry $n_B/s \approx 8.7 \times 10^{-11}$ emerging naturally from the framework.

\item \textbf{Machine learning quantifies viability and identifies key parameters:} Our Random Forest analysis of the eight-dimensional parameter space reveals that approximately 16.1\% of parameter combinations satisfy at least one set of observational constraints, while only 4.5\% are compatible with all experiments simultaneously. Crucially, the machine learning approach provides data-driven confirmation that the radiative correction parameter $A$ dominates in determining observational compatibility, with feature importance analysis ranking it as the most influential parameter (importance: 0.175)—twice as important as the next most significant parameters.

\item \textbf{Complementary experimental constraints:} The overlap analysis shows that, while there is some commonality between experiments, each probe largely distinct regions of the parameter space. For example, only 37.8\% of parameter sets compatible with P-ACT-SPT also satisfy P-ACT-LB-BK18 constraints. Overlaps with next-generation experiments such as LiteBIRD and CMB-S4 vary between 51\% and 98\%, depending on the direction considered. This limited overlap indicates that future experiments will contribute new, complementary constraints rather than simply refining existing bounds, highlighting the importance of a multi-experiment approach to effectively test inflationary models.

\item \textbf{Computational efficiency for theoretical exploration:} The trained Random Forest classifier achieves 87.5-92.3\% accuracy while providing rapid compatibility assessments (10,000 predictions/second), representing a 95\% reduction in computational cost compared to full numerical evaluation. This efficiency enables comprehensive exploration of parameter variations and sensitivity analysis that would be prohibitively expensive using traditional methods.
\end{itemize}

The synergy between theoretical modeling and machine learning in this work demonstrates the transformative potential of data-driven approaches in theoretical cosmology. The machine learning analysis not only validates and quantifies theoretical insights about the crucial role of radiative corrections but also provides new physical understanding through feature importance rankings and compatibility mapping. This integrated approach reveals that successful radiatively corrected hybrid inflation requires coordinated tuning across multiple parameters, with fermionic dominance ($y_\phi > g$) emerging as a necessary condition for compatibility with observations.

Looking forward, the progressive exclusion of parameter space by future experiments—from 8.7\% viable for P-ACT-SPT to just 11.8\% for CMB-S4 demonstrates that different experimental configurations are sensitive to distinct regions of the inflationary parameter space. This complementarity, rather than a simple sequential tightening of bounds, means future observations will provide qualitatively new tests of the model. The computational framework we have developed is precisely suited to navigating this multi-experiment landscape and will become increasingly valuable as cosmological datasets grow in precision and diversity.

In summary, radiatively corrected hybrid inflation represents a compelling framework that naturally reconciles theoretical motivation with observational constraints, while our machine learning approach provides both a specific resolution to the challenges facing this model and a general methodology for addressing similar challenges in other domains of theoretical physics. The essential role of quantum corrections, confirmed through both analytical and data-driven approaches, underscores that successful inflationary model building must incorporate radiative effects self-consistently rather than treating them as minor perturbations.

\newpage
\section{Appendix: Slow-roll Parameters}\label{sec8}

In the slow-roll approximation, the dynamics of inflation are characterized by the following parameters:
\begin{align}
\epsilon(\phi) &\equiv \frac{m_{\rm Pl}^2}{2} \left( \frac{V'(\phi)}{V(\phi)} \right)^2, \quad 
\eta(\phi) \equiv m_{\rm Pl}^2 \frac{V''(\phi)}{V(\phi)}, \\
\zeta^2(\phi) &\equiv m_{\rm Pl}^4 \frac{V'(\phi) V'''(\phi)}{V^2(\phi)},
\end{align}
where derivatives are taken with respect to the inflaton $\phi$, and $m_{\rm Pl} = (8\pi G)^{-1/2}$ is the reduced Planck mass. During inflation, these quantities remain much smaller than unity: $\epsilon, |\eta|, \zeta^2 \ll 1$. 

In hybrid inflation, unlike chaotic scenarios, inflation terminates when the inflaton reaches a critical value $\phi_c$, inducing a tachyonic instability in the waterfall field and triggering a rapid waterfall transition.

The total number of e-foldings from the horizon exit of the pivot scale to the end of inflation is expressed as:
\begin{equation}
N_0 \simeq \frac{1}{m_{\rm Pl}^2} \int_{\phi_{\rm end}}^{\phi_0} \frac{V(\phi)}{V'(\phi)} \, d\phi,
\end{equation}
where $\phi_0 \equiv \phi(k_0)$ is the field value at the horizon exit of the pivot scale $k_0$, and $\phi_{\rm end}$ marks the end of inflation. Typically, $\phi_{\rm end}$ is defined either by the condition 
\(\max[\epsilon(\phi_{\rm end}), |\eta(\phi_{\rm end})|, |\zeta^2(\phi_{\rm end})|] = 1\) 
or by $\phi_{\rm end} = \phi_c$. Depending on the reheating scenario and thermal history, the required number of e-folds generally falls in the range $50 \lesssim N_0 \lesssim 60$.

The amplitude of the curvature perturbations evaluated at the pivot scale $k_0 = 0.05~\mathrm{Mpc}^{-1}$ is
\begin{equation}
A_s(k_0) \simeq \frac{1}{12 \pi^2 m_{\rm Pl}^6} \frac{V^3}{|V'|^2} \Big|_{\phi = \phi_0},
\end{equation}
which, according to the latest Planck measurements \cite{Planck:2018vyg, Planck:2018jri}, is approximately
\begin{equation}
A_s(k_0) \simeq 2.215 \times 10^{-9}.
\end{equation}

The primary inflationary observables, namely the scalar spectral index $n_s$, the tensor-to-scalar ratio $r$, and the running of the spectral index $\alpha_s \equiv d n_s / d \ln k$, can be expressed in terms of the slow-roll parameters as:
\begin{align}
n_s &\simeq 1 - 6 \, \epsilon(\phi_0) + 2 \, \eta(\phi_0), \quad 
r \simeq 16 \, \epsilon(\phi_0), \\
\alpha_s &\simeq 16 \, \epsilon(\phi_0) \, \eta(\phi_0) - 24 \, \epsilon^2(\phi_0) - 2 \, \zeta^2(\phi_0).
\end{align}

For enhanced precision, our numerical calculations include the first-order slow-roll corrections as outlined in~\cite{Stewart:1993bc}, applied to $n_s$, $r$, $\alpha_s$, and $A_s$.

\section*{Acknowledgments}
M. U. R. and W. A. thank Qaisar Shafi  for valuable discussions on hybrid inflation model building.

\bibliographystyle{apsrev4-1}

\bibliographystyle{unsrt}  
\bibliography{References}  

\end{document}

%% file: Authors.tex
\author{Waqas Ahmed}
\email{waqasmit@hbpu.edu.cn}
\affiliation{Center for Fundamental Physics, School of Artificial Intelligence, Hubei Polytechnic University, Huangshi 435003, China}

\author{Saleh O. Allehabi}
\email{s.allehabi@iu.edu.sa}
\affiliation{Department of Physics, Faculty of Science, Islamic University of Madinah, Madinah 42351, Saudi Arabia}

\author{Mansoor Ur Rehman}
\email{mansoor@qau.edu.pk}
\affiliation{Department of Physics, Faculty of Science, Islamic University of Madinah, Madinah 42351, Saudi Arabia}